\documentclass[english,aps,prapplied,twocolumn,superscriptaddress,longbibliography]{revtex4-2}
\usepackage[urlcolor=blue, hyperindex, colorlinks, bookmarks=true]{hyperref}

\usepackage{graphicx}% Include figure files
\usepackage{dcolumn}% Align table columns on decimal point
\usepackage{bm}% bold math
\usepackage{amsmath,color,amssymb}
\usepackage[normalem]{ulem}
\usepackage{bm}
\usepackage{soul}
\usepackage[utf8]{inputenc}
\usepackage[T1]{fontenc}
\usepackage{mathptmx}
\usepackage{mathtools}
\usepackage{float}
\RequirePackage[normalem]{ulem} % strikethrough

\definecolor{new}{rgb}{.38,.6,.38}
\definecolor{old}{rgb}{1,0,0}
\definecolor{off}{rgb}{0,0,0}

\newcommand{\princeton}{\affiliation{Department of Physics, Princeton University, Princeton, New Jersey~08544, USA}}
\newcommand{\UCLA}{\affiliation{Department of Physics and Astronomy, University of California -- Los Angeles, Los Angeles, California 90095, USA}}
\newcommand{\CQSE}{\affiliation{Center for Quantum Science and Engineering, University of California -- Los Angeles, Los Angeles, California 90095, USA}}

\begin{document}
\title{Developing high-impedance superconducting resonators and on-chip filters for semiconductor quantum dot circuit quantum electrodynamics}

\author{X. Zhang}
\princeton
\author{Z. Zhu}  
\princeton
\author{N. P. Ong} 
\princeton
\author{J. R. Petta}
\UCLA
\CQSE

\begin{abstract}
Spin-photon coupling presents an enticing opportunity for the long-range coupling of spin qubits. The spin-photon coupling rate $g_{s}$ is proportional to the charge-photon coupling rate $g_{c}$. To move deeper into the strong coupling regime, $g_{c}$ can be enhanced by fabricating high-impedance cavities using high kinetic inductance films. Here we report dc transport and microwave response investigations of niobium nitride (NbN) films of different thicknesses. The kinetic inductance increases rapidly as the film thickness is reduced below 50 nm and for 15~nm NbN films we measure a sheet kinetic inductance $L_{k,S}$ = 41.2 pH/$\Box$. As an application of the high kinetic inductance films, we fabricate compact LC filters that are commonly used to reduce microwave leakage in circuit quantum electrodynamics (cQED) devices. These filters feature up to 60 dB of attenuation near typical cavity resonance frequencies $f_c$ = 8 GHz. %NbN thin films, and the filters they enable, have the potential to move quantum dot cQED devices deeper into the strong coupling regime.
\end{abstract}

\maketitle

\section{Introduction}

In recent years, quantum computing has been a field of fast-growing interest and intense research. In the quest for a large-scale quantum computing architecture, individual processing units can be interconnected via a coherent mediator, allowing entanglement to be spread across the quantum circuit. Many different schemes for short-range nearest-neighbor entanglement have been demonstrated and perfected \cite{petta2005coherent,mortemousque2021coherent,watson2018programmable,zajac2018resonantly}. However, long-range coupling between distant qubits has remained a challenge. To date, some approaches to achieve long-range qubit entanglement include coherently shuttling electrons using gate-voltage pulses \cite{baart2016single,mills2019shuttling} or surface acoustic waves \cite{de2005modulation,stotz2005coherent,hermelin2011electrons,mcneil2011demand}, coupling spin qubits with a much larger coupling quantum dot (QD) \cite{malinowski2019fast}, or using a microwave cavity as a mediator \cite{mi2018coherent, samkharadze2018strong,landig2018coherent,borjans2020resonant,borjans2020split}.

Strong spin-photon coupling has been achieved in circuit quantum electrodynamics (cQED) device architectures by isolating a single electron in a Si double quantum dot (DQD) and utilizing the field gradient from a nearby micromagnet to engineer a synthetic spin-orbit interaction \cite{mi2018coherent,samkharadze2018strong}. Exchange-only spin qubits fabricated in GaAs have also reached the strong coupling regime \cite{landig2018coherent}. Resonant coupling of two spin qubits separated by $\sim$ 4 mm has been demonstrated \cite{borjans2020resonant} and attempts have been made to extend cavity-mediated spin-spin coupling into the dispersive regime \cite{harvey2021circuit}. However, a larger spin-photon coupling rate $g_{s}$ is needed to achieve high-fidelity remote gates between spatially separated spins. 

The spin-photon coupling rate is set by the electric dipole moment of the electron trapped in the DQD, the cavity electric field, and the strength of the effective spin-orbit interaction. The cavity field is a function of the characteristic impedance of the resonator $Z_{C}$. Therefore, an enhancement in $g_{s}$ can be achieved by increasing $Z_{C}$, as $g_{s}\propto g_{c}\propto \sqrt{Z_{C}}$. The kinetic inductance can be greatly increased using large arrays of superconducting quantum interference devices (SQUIDs) \cite{stockklauser2017strong}. However, for single spin qubits, the magnetic fields required to Zeeman split the spin states interferes with the operation of SQUID arrays.

Another way to increase characteristic impedance of the resonator is to fabricate it from a high kinetic inductance superconducting film. The resonator impedance is given by $Z_{C}=\sqrt{\frac{L}{C}}$, where $L$ and $C$ are the characteristic inductance and capacitance. The inductance $L = L_{m}+L_{k}$, can be separated into magnetic, $L_{m}$, and kinetic, $L_{k}$, components. In past experiments, the magnetic inductance has been modestly increased using narrow Nb stripline resonators with an estimated impedance $Z_{C}$ $\sim$ 300 $\Omega$  \cite{mi2018coherent}. For some superconducting materials, $L_{k}$ is dominant compared with $L_{m}$ \cite{zmuidzinas2012superconducting}. Therefore, to tune $Z_{C}$ to a greater extent, high kinetic inductance materials are desirable. Since $L_{k}$ is directly proportional to the ratio of the resonator length $l$ to the cross-sectional area of the superconducting film \cite{tinkham2004introduction}, an effective approach is to fabricate a narrow resonator from a thin film of a high $L_{k}$ material \cite{harvey2020chip, harvey2021circuit}.  
      
In this Article, we study high $L_{k}$ NbN thin films and present two different approaches for extracting the sheet kinetic inductance $L_{k,S}$. We begin with dc resistivity measurements across the superconducting transition and compute $L_{k,S}$ over a large range of film thicknesses. We next characterize the microwave response of quarter-wavelength ($\lambda$/4) hanger-style resonators fabricated from NbN. To investigate the thickness dependence of $L_{k,S}$, we measure a variety of resonators, and from the extracted resonance frequencies we determine the kinetic inductance contribution to the total inductance. The kinetic inductance values $L_{k,S}$ extracted from dc and microwave measurements are in good agreement. Lastly, we utilize thin NbN films to fabricate compact LC filters that are compatible with QD-cQED devices. These filters feature up to 60~dB of attenuation near typical cavity resonance frequencies $f_c$~=~8~GHz and have a footprint that is significantly smaller than our previous design \cite{mi2018coherent}.

\section{Niobium nitride film deposition and dc resistivity measurements}
\begin{figure*}[tbp]
	\centering
	\includegraphics[width=2\columnwidth]{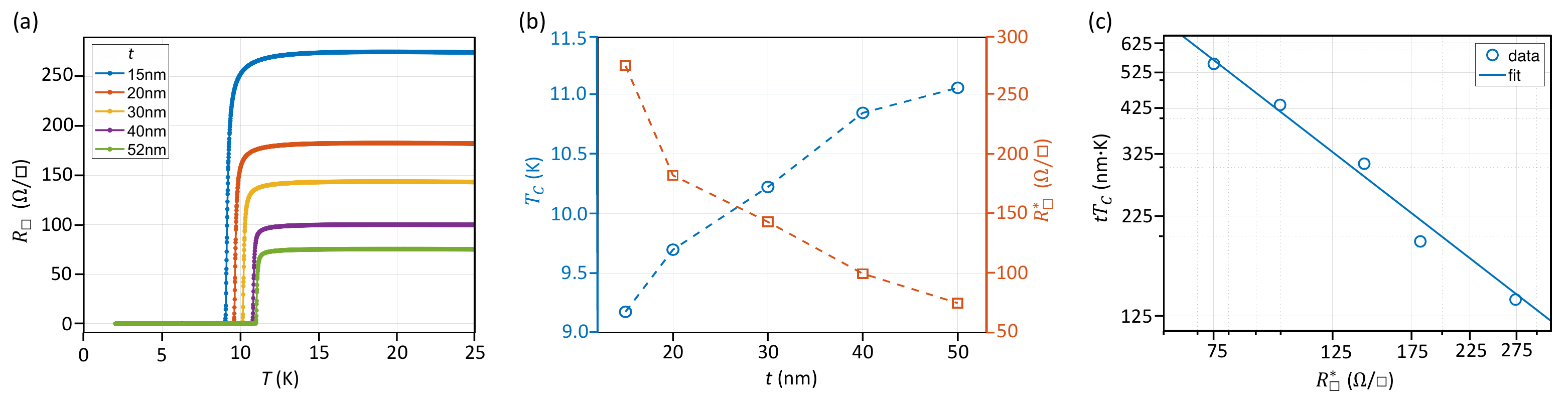}
	\caption{(a) Temperature dependence of the sheet resistance $R_{\Box}$ for NbN thin films of different thicknesses $t$ measured near the superconducting transition. (b) Critical temperature $T_{\rm C}$ and normal state sheet resistance $R_{\Box}^*$ as a function of $t$. The dashed lines are guides to the eye. (c) $tT_{C}$ plotted as a function of $R^*_{\Box}$. The blue solid line is a fit to the universal scaling relation $tT_{C} = A{R^*_{\Box}}^{-b}$, yielding the power-law exponent $b$ = 1.1 $\pm$ 0.3.}
	\label{fig:1}
\end{figure*}

We begin by characterizing the sheet resistance and critical temperature $T_C$ of unpatterned superconducting NbN thin films using dc transport measurements. From the Bardeen-Cooper-Schrieffer (BCS) theory of superconductivity, these quantities can be used to estimate the kinetic inductance \cite{zmuidzinas2012superconducting, bardeen1957theory}. 

We use an AJA dc-magnetron sputtering system with a high-purity Nb target to deposit the NbN films. Taking the critical temperature $T_C$ as a metric, we optimize the deposition parameters, including gas flow rates, target power, and deposition pressure. With a base pressure below 2$\times$10$^{-9}$~Torr, we initiate the deposition process by introducing N$_2$ and Ar into the system at flow rates of 6 sccm and 60 sccm to condition the chamber. We then bias the Nb target with a power of 400 W to ignite a plasma. The plasma of ions dislodges Nb atoms from the target, and the Nb atoms react with N$_2$ to form NbN, which then deposits onto the substrate. The chamber pressure is held at 10 mTorr during deposition. We deposited five NbN thin films of various thicknesses, with the substrate held at room temperature, utilizing a deposition rate of approximately 12~\r{A}/s.

The thin films are characterized in a physical property measurement system (PPMS). We measure the sheet resistance $R_{\Box}$ as a function of temperature $T$ using conventional 4-probe transport techniques. As shown in Fig.~\ref{fig:1}(a), all films exhibit sharp superconducting transitions below $T$ = 12 K. From each cooldown curve, we extract the superconducting critical temperature $T_C$ and the normal state sheet resistance just before the superconducting transition $R_{\Box}^*$.
 $T_C$ is defined as the temperature at which $R_{\Box}$ drops to half of its normal state value. $T_C$ and $R_{\Box}^*$ are plotted as a function of film thickness $t$ in Fig.~\ref{fig:1}(b). As the film thickness $t$ decreases, $R_{\Box}^*$ increases, while $T_C$ monotonically decreases. The general trends observed in the temperature dependence of $T_{C}$ and $R_{\Box}^*$ are in agreement with prior work \cite{chockalingam2008superconducting,hazra2016superconducting,shibalov2021multistep}. In Fig.~\ref{fig:1}(c), we plot $tT_{C}$ as a function of $R_{\Box}^*$. The data are fit to the scaling relation $tT_{C} = A{R^*_{\Box}}^{-b}$ \cite{ivry2014universal}. We extract the exponent $b = 1.1\pm 0.3$, which agrees with $b\approx 0.9\sim 1.1$ widely observed in high-quality superconducting thin films \cite{ivry2014universal,faverzani2020characterization,amin2021cmos}. 

Using the extracted $T_{C}$ and $R_{\Box}^*$, we estimate the sheet kinetic inductance using the formula \cite{zmuidzinas2012superconducting}
\begin{equation}
L_{k,S} = \frac{\hbar R_{\Box}^*}{\pi\Delta_{0}},
\end{equation}
where $\hbar$ is the reduced Planck's constant and $\Delta_{0}$ is the zero temperature superconducting gap. Assuming the NbN thin films obey BCS theory, $\Delta_{0} = 1.76\times k_{B}T_{C}$, where $k_{B}$ is Boltzmann's constant \cite{bardeen1957theory}. The extracted values of $T_{C}$, $R_{\Box}^*$, and the evaluated results of $L_{k,S}$, are given in Table~\ref{tbl:1}. As $t$ decreases from 52 nm to 15 nm, we observe a large increase in $R_{\Box}^*$ from 75.1~$\Omega /\Box$ to 273.9~$\Omega /\Box$ and a small decrease in $T_{C}$ from 11.1~K to 9.2~K, resulting in a significant increase in $L_{k, S}$ from 9.4~pH/$\Box$ to 41.2~pH/$\Box$. These estimated kinetic inductance values allow for the efficient design of microwave cavities. 

\begin{table}[t]
\begin{tabular}{|c|c|c|c|}
\hline

$\quad t\: (\rm{nm})\quad$ 	& $\quad {T_C}\: (\rm{K})\quad$ 	& $\quad R^*_{\Box}\: (\Omega/\Box)\quad$   & $\quad L_{k, S}\: (\rm{pH}/\Box)\quad$ \\ 

\hline
$15$	    & $9.2$             &$273.9$           &$41.2$\\
\hline
$20$ 		& $9.7$             &$182.1$           &$25.9$\\
\hline
$30$ 		& $10.2$            &$143.2$           &$19.3$\\
\hline
$40$ 		& $10.8$            &$99.8$            &$12.7$\\
\hline
$52$ 		& $11.1$            &$75.1$            &$9.4$\\
\hline
\end{tabular}
\caption{\label{tbl:1} Critical temperature $T_{C}$, normal state sheet resistance $R^*_{\Box}$, and sheet kinetic inductance $L_{k,S}$ for various film thicknesses $t$.}
\end{table}

\section{Microwave resonator characterization of the kinetic inductance}

\begin{figure*}[t]
	\centering
	\includegraphics[width=2\columnwidth]{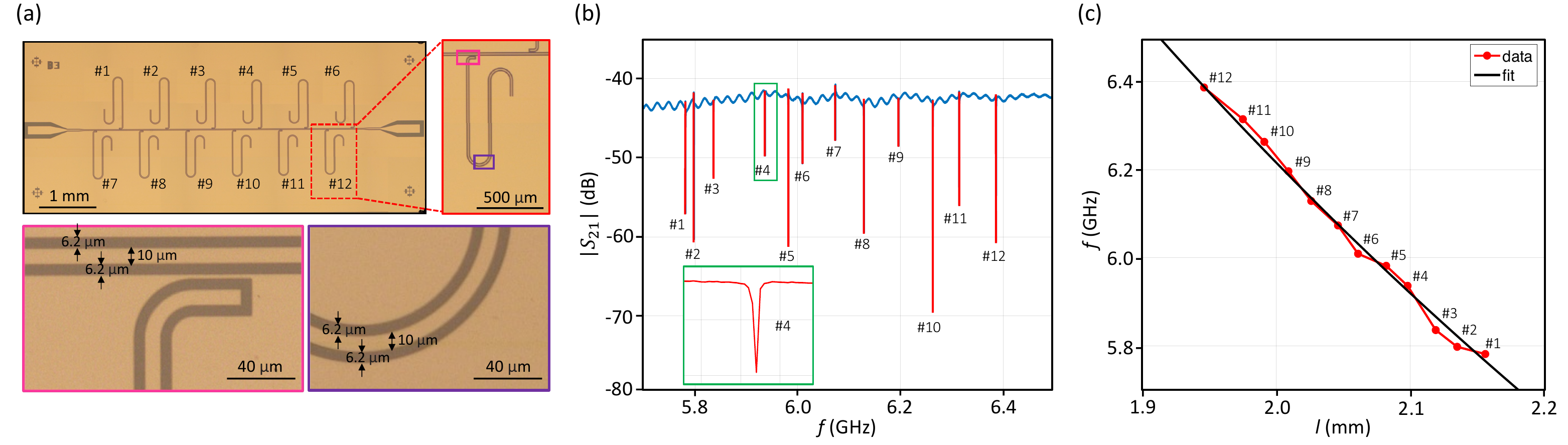}
	\caption{Microwave characterization of hanger-style resonators. (a) Optical images of samples containing 12 $\lambda$/4  resonators of varying length $l$ that are coupled to a central transmission line. Insets: Zoomed-in images showing resonator dimensions. (b) Microwave transmission $|S_{21}(f)|$ through a sample made from a 30 nm NbN film. The 12 resonances are highlighted in red. Inset: Zoom-in of $|S_{21}(f)|$ near the resonance at $f$~=~5.9 GHz. (c) Extracted resonance frequencies plotted as a function of $l$, showing the expected $1/l$ dependence. A linear fit to the data (solid line) using a transmission line model yields a sheet kinetic inductance $L_{k,S}$ = 20.9 pH/$\Box$.}
	\label{fig:2}
\end{figure*}
%\vspace{-0.5cm}

Using the kinetic inductance values obtained in Sec.~II, we design microwave cavities with resonance frequencies ranging from 5.8 -- 6.4 GHz. We fabricate an array of $\lambda$/4 cavities that are coupled to a transmission line. The side-coupled cavities allow for the efficient extraction of resonance frequencies and quality factors \cite{tinkham2004introduction,yoshida1995evaluation}. By fabricating resonators with different dimensions, we can quantitatively extract the magnetic and kinetic contributions to the total inductance. 

As shown in Fig.~\ref{fig:2}(a), a typical resonator chip contains 12 $\lambda$/4 resonators of varying length $l$, each capacitively coupled to a central transmission line. The sample is made from a 30~nm NbN thin film sputtered on a high-resistivity Si substrate, and the resonators are of the coplanar waveguide (CPW) design with a central conductor width $a$ = 10 $\mu$m and a gap width $s$ = 6.2 $\mu$m. We use a network analyzer to characterize the microwave response of the sample, $|S_{21}(f)|$, in a dilution refrigerator with a lattice temperature of approximately 10 mK. Each resonance results in a Lorentzian dip in $|S_{21}(f)|$. In Fig.~\ref{fig:2}(b), 12 resonances are observed (highlighted in red), corresponding to the 12 different resonator lengths. 

In Fig.~\ref{fig:2}(c), we plot the extracted resonance frequencies of all 12 resonators and observe the expected relationship $f\propto 1/l$. For a $\lambda$/4 resonator, the resonator length can be written as 
\begin{equation}
l = \lambda/4 = \frac{v_{\rm eff}}{4f_0},
\end{equation}
where $\lambda$ is the wavelength and $f_0$ is the resonance frequency. For non-magnetic substrates, the propagation speed is $v_{\rm eff} = \frac{c}{\sqrt{\mu_{\rm eff}\epsilon_{\rm eff}}} \approx \frac{c}{\sqrt{\epsilon_{\rm eff}}}$, where $c$ is the speed of light in vacuum, $\mu_{\rm eff}$ is the relative magnetic permeability, and $\epsilon_{\rm eff}$ is the relative dielectric constant, defined as
\begin{equation}
\epsilon_{\rm eff} = \frac{1+\epsilon_r\tilde{K}}{1+\tilde{K}}.
\end{equation} The value of $\epsilon_{\rm eff}$ is determined by the geometry of the CPW resonator and the relative dielectric constant of the substrate $\epsilon_r$ = 11.7, with $\tilde{K}$ given by 
\begin{equation}
\tilde{K} = \frac{K(k')K(k_3)}{K(k)K(k_3')}.    
\end{equation}
Here $K$ is the complete elliptical integral of the first kind, with $k = \frac{a}{b}$, $k_3 = \frac{\tanh{(\frac{\pi a}{4h}})}{\tanh{(\frac{\pi b}{4h})}}$, $k' = \sqrt{1-k^2}$, $k_3' = \sqrt{1-k_3^2}$, and $b = a + 2s$, where $a$ is the center conductor width, $s$ is the gap width, and $h$ is the substrate thickness \cite{schuster2007circuit, mi2018circuit}.

In a CPW geometry, the magnetic inductance per unit length is given by \cite{yoshida1995evaluation}:
\begin{equation}
L_m = \frac{\mu}{4}\frac{K(k')}{K(k)},
\end{equation}
where $\mu$ is the magnetic permeability of the substrate. In the case of a high-$L_k$ resonator, the kinetic inductance must be included to obtain the total inductance, $L = L_m + L_k$. With $f \propto 1/\sqrt{L}$, we can fit to the data to extract $L_{k,S}$ = 20.9 pH/$\Box$ [solid line in Fig.~\ref{fig:2}(c)], which is close to the estimate obtained from the PPMS data. As the magnetic inductance only depends on the CPW geometry, we use the resonator dimensions to calculate the sheet magnetic inductance $L_{m,S}$ = 4.4 pH/$\Box$.

\begin{figure}[tbp]
	\centering
	\includegraphics[width=\columnwidth]{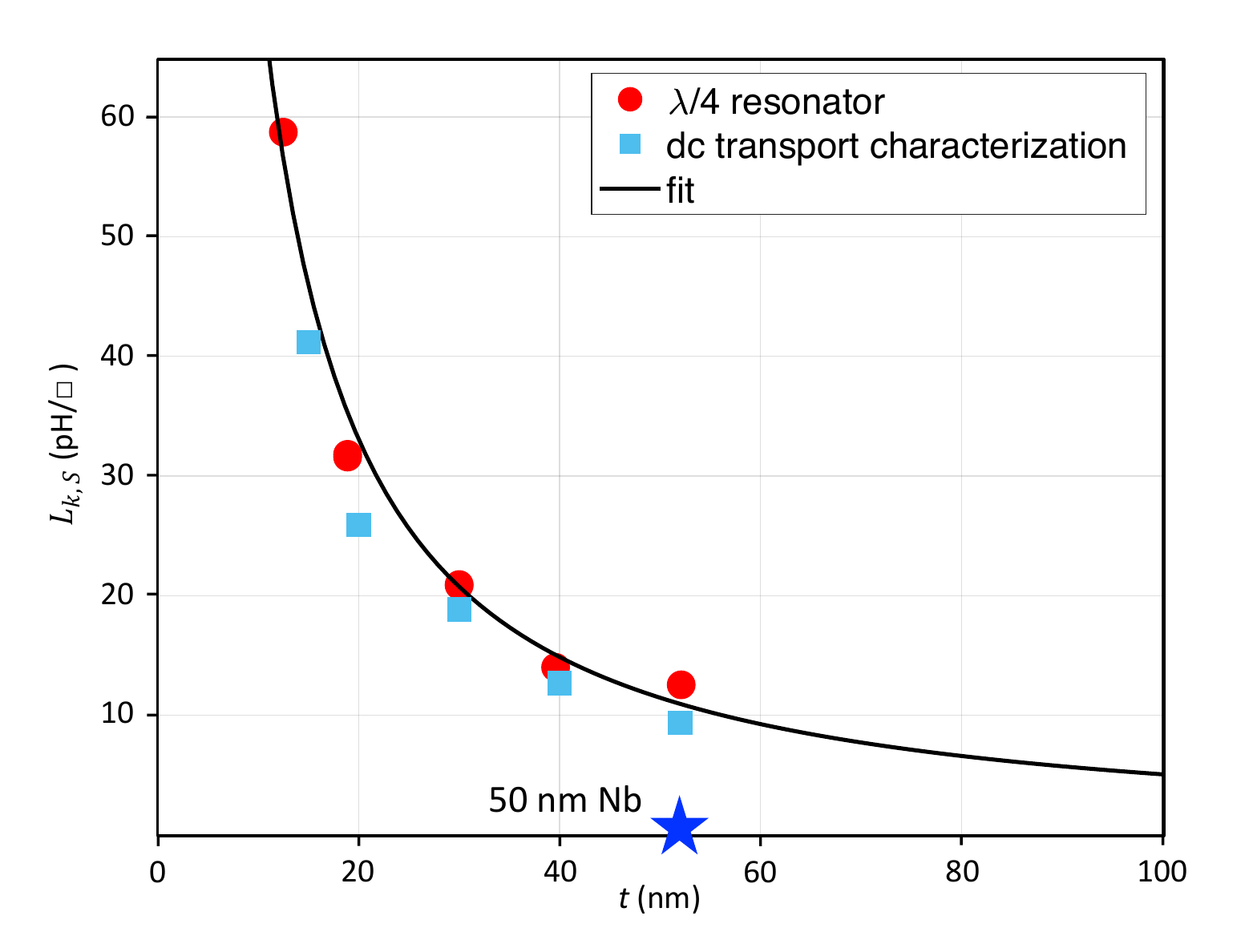}
	\caption{Thickness dependence of the sheet kinetic inductance $L_{k,S}$. The data are extracted from the dc characterization of the thin film resistivity (Sec.~II) and the microwave response of $\lambda/4$ resonators (Sec.~III). The blue star indicates the much smaller $L_{k,S}$ = 0.5 pH/$\Box$ of a standard 50 nm Nb film.}
	\label{fig:3}
\end{figure}

To investigate the thickness dependence of $L_{k,S}$, we fabricate a variety of hanger-style resonators using NbN thin films of different thicknesses. For each film thickness, we determine $L_{k,S}$ by fitting to the measured resonance frequencies, as described in the previous paragraph. Figure~ \ref{fig:3} combines results from dc resistivity measurements and microwave resonator characterization. The values of $L_{k,S}$ extracted from microwave measurements are in good agreement with the estimates obtained from dc transport measurements. The kinetic inductance of the NbN films is significant. For a direct comparison, the kinetic inductance extracted from a set of resonators fabricated on a conventional 50 nm thick Nb film is only $L_{k,S}$ = 0.5 pH/$\Box$. 

As expected from theory, $L_{k,S}$ rapidly increases as $t$ is reduced below 50 nm, with the highest value $L_{k,S}$ = 58.7 pH/$\Box$ obtained from hanger resonators made from a 12.5 nm film. Our past cQED devices utilized 50 nm Nb films and the largest coupling rates measured were $g_{c} \approx$ 58 MHz in Ref.~\cite{borjans2020split} and $g_{c(s)} \approx$ 40 (11) MHz in Ref.~\cite{borjans2020resonant}. To estimate the potential enhancement in $g_{c(s)}$ by switching to 15 nm NbN with $L_{k,S}$~=~41.2 pH/$\Box$, we compare the total inductances assuming the same $\lambda$/2 resonator design using the two different materials. The $\lambda$/2 resonators on our current QD-cQED chips have a center pin width $a$ = 0.75 $\mu$m and gap width $s$ = 19.63 $\mu$m, which yields a sheet magnetic inductance $L_{m,S}$ = 0.85 pH/$\Box$. With 15 nm NbN thick films, the magnetic inductance is only 2$\%$ of the total inductance. Since $g_{c(s)}\propto \sqrt{Z_{C}}\propto L^{1/4}$, switching from 50 nm Nb to 15 nm NbN is expected to result in a 5.6$\times$ increase in $Z_{C}$, and hence a 2.4$\times$ increase in $g_{c(s)}$.

\section{LC microwave filters for quantum dot circuit quantum electrodynamics}

\begin{figure}[t]
	\centering
	\includegraphics[width=\columnwidth]{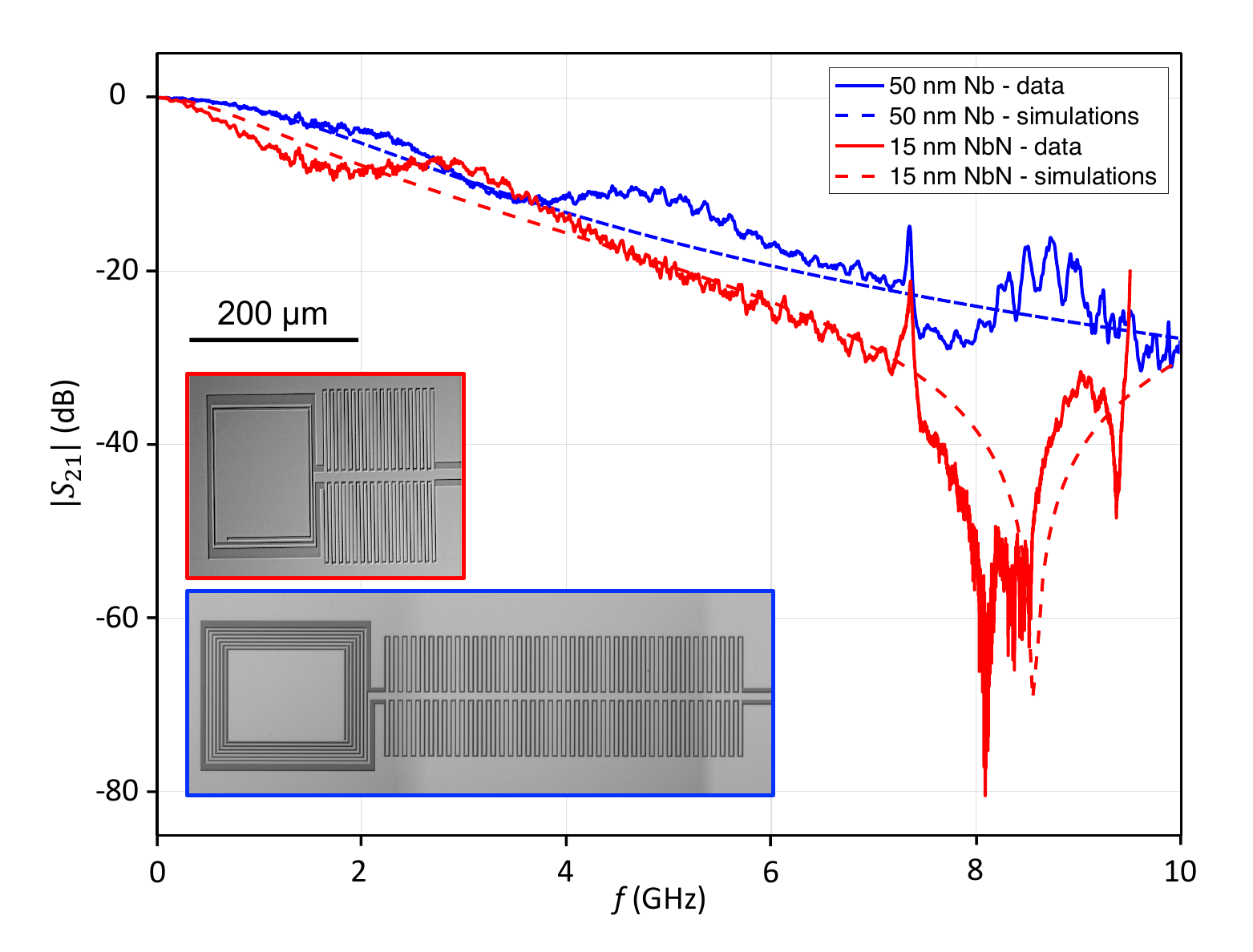}
	\caption{Measurements of $|S_{21}(f)|$ for two LC filters (solid lines) and the simulated response (dashed lines). The 15 nm NbN filter (red) gives a substantial attenuation of 60 dB around 8 GHz. In comparison, we measure 24 dB of attenuation using a $\sim$2$\times$ longer filter made from a 50 nm Nb film (blue). Insets: Optical images of filters fabricated from a 15 nm NbN film (outlined in red) and a 50 nm Nb film (outlined in blue).  
}
	\label{fig:4}
\end{figure}

We now utilize high kinetic inductance NbN films to fabricate compact microwave filters. Measurements of early QD-cQED devices show significant microwave leakage through the wiring used to dc bias the QD gate electrodes~\cite{petersson2012circuit}. The resulting photon leakage pathways can lower the cavity quality factor significantly. To suppress photon leakage, on-chip low-pass LC filters are often connected in series with each gate line \cite{mi2017circuit}. Each LC filter consists of a spiral inductor and an interdigitated capacitor (see Fig.~\ref{fig:4} insets). We simulate different designs using the Sonnet EM simulation package and fabricate Nb and NbN filters to evaluate their performance. 

Figure~\ref{fig:4} shows the microwave transmission $|S_{21}(f)|$ through two types of filters that are cooled to $T$ $\sim$ 700 mK. The filter fabricated from a standard 50 nm thick Nb film provides $\sim$24 dB of attenuation near a typical cavity frequency of 8 GHz \cite{mi2017circuit}. In contrast, a filter fabricated using a 15 nm thick NbN film is approximately 2$\times$ shorter and provides a much larger attenuation of 60 dB near 8 GHz. The experimental data are in good agreement with Sonnet simulations. These measurements show that the utilization of NbN films can significantly reduce filter sizes and provide a higher degree of attenuation to reduce cavity losses. 

\section{Conclusion}

In summary, we have investigated the thickness dependence of the sheet kinetic inductance of sputtered NbN thin films using a combination of dc transport and microwave measurements. To mitigate microwave leakage through the QD dc gate lines, we tested different LC filter designs and demonstrated 60 dB of attenuation near typical cavity resonance frequencies. Owing to the high $L_{k}$ of NbN, we can now achieve much stronger on-chip filtering with a smaller filter footprint. The high kinetic inductance $L_{k,S}$ = 41.2 pH/$\Box$ of the 15 nm NbN film provides an opportunity to further enhance $g_c$ and $g_s$, thereby moving QD-cQED systems further into the strong coupling regime.

\begin{acknowledgements}
Supported by Army Research Office grants W911NF-15-1-0149 and W911NF-23-1-0104. Devices were fabricated in the Princeton University Quantum Device Nanofabrication Laboratory, and in part in the Singh Center for Nanotechnology, which is supported by the NSF National Nanotechnology Coordinated Infrastructure Program under grant NNCI-2025608. NPO acknowledges support from the U.S. Department of Energy (DE-SC0017863).
\end{acknowledgements}

\newpage

\bibliography{Zhang_HighZ_v2}

\end{document}